\documentclass[useAMS,usenatbib,usedcolumn]{mn2e}
\def\HI{\ifmmode{\rm HI}\else{H\/{\sc i}}\fi}

\def\lsun{\ifmmode{{\mathrm L}_{\odot}}\else{L$_{\odot}$}\fi} 

\def\deg{\hbox{$^\circ$}}

\def\magasas{\nobreak\mbox{$\;$mag$\,$arcsec$^{-2}$}}
\def\msun{\ifmmode{{\mathrm M}_{\odot}}\else{M$_{\odot}$}\fi} 
\def\msunpc2{\ifmmode{{\mathrm M}_{\odot} \, {\mathrm{pc}}^{-2}}\else{M$_{\odot} \, {\mathrm {pc}}^{-2}$}\fi}

\def\kms{\ifmmode{{\mathrm{km \, s^{-1}}}}\else{${\mathrm{km \, s^{-1}}}$}\fi}

\def\ga{\mathrel{\mathchoice {\vcenter{\offinterlineskip\halign{\hfil
$\displaystyle##$\hfil\cr>\cr\sim\cr}}}
{\vcenter{\offinterlineskip\halign{\hfil$\textstyle##$\hfil\cr
>\cr\sim\cr}}}
{\vcenter{\offinterlineskip\halign{\hfil$\scriptstyle##$\hfil\cr
>\cr\sim\cr}}}
{\vcenter{\offinterlineskip\halign{\hfil$\scriptscriptstyle##$\hfil\cr
>\cr\sim\cr}}}}}

\usepackage[figuresright]{rotating}
\usepackage{lscape}
\usepackage{graphics}
\usepackage{epsfig}
\usepackage{multirow}
\usepackage{bigdelim}
\usepackage{bigstrut}

\defcitealias{Noordermeer07b}{N07}
\defcitealias{Tully96}{T96}
\defcitealias{Verheijen01a}{V01}
\defcitealias{Spekkens06}{S06}

\title[Massive galaxies on the Tully-Fisher relation]{The high mass end of the
  Tully-Fisher relation} 
\author[E.~Noordermeer \& M.~A.~W.~Verheijen]
  {E.~Noordermeer,$^{1,2}$\thanks{email:edo.noordermeer@nottingham.ac.uk} 
  and M.~A.~W.~Verheijen$^2$ \\ 
  $^1$University of Nottingham, School of Physics and Astronomy, University
  Park, NG7 2RD Nottingham, UK \\
  $^2$Kapteyn Astronomical Institute, University of Groningen, PO Box 800,
  9700 AV Groningen, The Netherlands}

\begin{document}

\date{accepted version, 20-08-2007}

\maketitle

\begin{abstract}
  We study the location of massive disk galaxies on the Tully-Fisher relation.
  Using a combination of K-band photometry and high-quality rotation curves,
  we show that in traditional formulations of the TF relation (using the width
  of the global \HI\ profile or the maximum rotation velocity), galaxies with
  rotation velocities larger than $200 \, {\mathrm {km \, s^{-1}}}$ lie
  systematically to the right of the relation defined by less massive systems,
  causing a characteristic `kink' in the relations. Massive, early-type disk
  galaxies in particular have a large offset, up to 1.5 magnitudes, from the 
  main relation defined by less massive and later-type spirals.\\ 
  The presence of a change in slope at the high-mass end of the Tully-Fisher
  relation has important consequences for the use of the Tully-Fisher relation
  as a tool for estimating distances to galaxies or for probing galaxy
  evolution. In particular, the luminosity evolution of massive galaxies since
  $z \approx 1$ may have been significantly larger than estimated in several
  recent studies.\\ 
  We also show that many of the galaxies with the largest offsets have
  declining rotation curves and that the change in slope largely disappears
  when we use the asymptotic rotation velocity as kinematic parameter. The
  remaining deviations from linearity can be removed when we simultaneously
  use the total baryonic mass (stars + gas) instead of the optical or
  near-infrared luminosity. Our results strengthen the view that the
  Tully-Fisher relation fundamentally links the mass of dark matter haloes
  with the total baryonic mass embedded in them.
\end{abstract}


%%%%%%%%%%%%%%%%%%%%%%%%%%%%%%%%%%%%%%%%%%%%%%%%%%%%%%%%%%%%%%%%%%%%%%%%%%%%%%%
%                                                                             %
%  1. Introduction                                                            %
%  \label{sec:introduction}                                                   %
%                                                                             %
%%%%%%%%%%%%%%%%%%%%%%%%%%%%%%%%%%%%%%%%%%%%%%%%%%%%%%%%%%%%%%%%%%%%%%%%%%%%%%%
\section{Introduction}
\label{sec:introduction}
The notion of a tight correlation between absolute luminosities of 
spiral galaxies and their rotational velocities has been with us for 
about thirty years now \citep{Tully77}.
This `Tully-Fisher' (TF) relation has been confirmed to hold over many decades 
in luminosity \citep{Courteau97, McGaugh00, Verheijen01a} and in different
galaxy environments \citep{Giovanelli97, Willick99}.  
Barred and unbarred galaxies share the same TF relation as well
\citep{Courteau03}, as do high and low surface brightness galaxies
\citep{Sprayberry95, Zwaan95}.
Although spirals of different morphological types do follow different TF
relations \citep{Roberts78, Rubin85, Hinz01, Mathieu02, Russell04}, these
offsets disappear almost entirely when using near-infrared photometry, rather
than optical luminosities \citep{Aaronson83, Peletier93}, indicating that most
of the differences at bluer wavelengths can be attributed to variations in
star formation history along the Hubble sequence and the resulting differences
in stellar populations.
Similarly, when dynamical models are used to derive circular velocities for
elliptical galaxies, it is found that the latter follow a similar TF relation
to spirals, but offset to lower luminosities. This offset can again be 
explained from differences in the stellar mass-to-light ratios
\citep{Gerhard01, DeRijcke07}. 

The Tully-Fisher relation has become one of the most widely used
relations in extragalactic astronomy. 
It has been commonly used as a powerful tool to estimate distances to
galaxies \citep[e.g.][and references therein]{Tully00}. 
As a statistical correlation between fundamental properties of spiral
galaxies, it has also been used to constrain numerical simulations of
galaxy formation \citep{Dalcanton97, Navarro00, Bullock01, Pizagno06,
Portinari07}, probe galaxy evolution on cosmological timescales \citep{Vogt97,
Ziegler02, Milvang-Jensen03, Bohm04, Bamford06, Weiner06, Chiu07}, or study
the structure and stellar populations of nearby galaxies
\citep{Franx92, Courteau99, Bell01}.  

Traditionally, the Tully-Fisher relation was studied using the width $W$ of a
galaxy's global neutral hydrogen line profile as a probe for its rotational
velocity \citep{Tully77}. 
With larger samples of spatially resolved kinematics (rotation curves)
becoming available, the rotation velocities can be directly measured and
recent studies tend to use the maximum rotation velocity ($V_{\mathrm {max}}$)
instead.   
This subtle shift has the added advantage that the TF relation can be
measured at cosmological distances, using optical spectroscopic observations
of high-redshift galaxies (see references above). 
 
Most spiral galaxies have flat rotation curves and for these systems, both
kinematic measurements ($W$ and $V_{\mathrm {max}}$) generally yield nearly
equivalent velocities.   
However, this is not the case for all galaxies. 
In galaxies whose rotation curves are rising or declining at large radii, it
is not clear a priori which kinematic parameter yields the best results. 
\citet[][hereafter V01]{Verheijen01a} presented an extensive study, based on a
sample of 31 galaxies in the Ursa-Major cluster with well-defined rotation
curves and K-band photometry, of the influence of the shape of a rotation
curve on the location of a galaxy on the Tully-Fisher relation.
Galaxies with declining rotation curves were found to lie systematically on
the high-velocity side, when using the maximum rotation velocity $V_{\mathrm
{max}}$ or the width of the \HI\ profile.   
Since these are typically bright systems with high rotation velocities, this
result suggests that there may be a change in slope in the TF relation at the
high-luminosity end, a possibility also hinted at by \citet{Peletier93}. 
Clearly, the existence of a `kink' in the Tully-Fisher relation has
important consequences and, if not corrected for, will lead to systematic
biases when deriving cosmological distances or probing galaxy evolution.  

However, \citetalias{Verheijen01a} also showed that the systematic
offset at the high mass end disappeared when, instead of $V_{\mathrm {max}}$,
the asymptotic rotation velocity $V_{\mathrm {asymp}}$ in the outer, flat
parts of the rotation curve was used.  
Bearing in mind that the asymptotic rotation velocities are determined 
by the dark matter haloes, \citetalias{Verheijen01a} interpreted this result
as a strong indication that the TF relation fundamentally links the total
baryonic content of (disk) galaxies with their dark matter haloes, and that as
far as this relation is concerned, there is no fundamental difference between
low  and high mass galaxies. 

Unfortunately, the sample of \citetalias{Verheijen01a} only contained a small
number of galaxies with declining rotation curves and the declines were
modest.
Thus, the change in slope of the high-luminosity end of the TF relation when
going from $V_{\mathrm {max}}$ to $V_{\mathrm {asymp}}$ was small and his
results only gave a first indication for how to interpret the curvature at the
high-mass end of the TF relation in terms of the kinematics of the galaxies. 
Recently, however, a sample of high-quality rotation curves for
high-luminosity early-type disk galaxies (S0 -- Sab) has become available
\citep[][hereafter N07]{Noordermeer07b}.  
This sample contains many strongly declining rotation curves where the
difference between the maximum and asymptotic rotation velocities is much
larger than in Verheijen's galaxies.  
In this paper therefore, we combine Noordermeer's sample with the data for the
galaxies with flat and declining rotation curves from
\citetalias{Verheijen01a}, with the aim of further investigating the change in
slope at the bright end of the TF relation and its relation with the shapes of
rotation curves.  
We also include the recent sample of rotation curves from \citet[][hereafter
S06]{Spekkens06}
in our analysis; these are all massive late-type spiral galaxies with flat
rotation curves and provide a valuable benchmark at the high-luminosity end of
the TF relation.

The outline of the remainder of this paper is as follows. 
In section~\ref{sec:sample+data}, we describe the selection of galaxies from
the three different samples for the present study, together with the
photometric and kinematic ingredients for the TF analysis. 
In section~\ref{sec:TFrelations}, we present the Tully-Fisher relations and
show that there is indeed a change in slope around $200 \, {\mathrm {km \,
s^{-1}}}$ in the relations with $W$ and $V_{\mathrm {max}}$.
We also show that this `kink' is largely, but not entirely, removed when we
use the asymptotic rotation velocity $V_{\mathrm {asymp}}$. 
In section~\ref{sec:discussion}, we search for possible explanations for the
observed change in slope and show that, when we consider the total baryonic
mass (stars + gas) in our galaxies, an entirely linear relation is recovered. 
Finally, we discuss some implications of our results in
section~\ref{sec:conclusions}.

%%%%%%%%%%%%%%%%%%%%%%%%%%%%%%%%%%%%%%%%%%%%%%%%%%%%%%%%%%%%%%%%%%%%%%%%%%%%%%%
%                                                                             %
%  2. Sample selection and observational data for the TF analysis             %
%  \label{sec:photdata}                                                       %
%                                                                             %
%%%%%%%%%%%%%%%%%%%%%%%%%%%%%%%%%%%%%%%%%%%%%%%%%%%%%%%%%%%%%%%%%%%%%%%%%%%%%%%
\section{Sample selection and observational data}
\label{sec:sample+data}

\subsection{Sample}
\label{subsec:sample}
We combine data from the three separate studies by \citet[][V01]{Verheijen01a},
\citet[][S06]{Spekkens06} and \citet[][N07]{Noordermeer07b}. 
The main criteria for inclusion of galaxies from these sub-samples in the
present study is that their rotation curves are of good quality and that the
maximum and asymptotic rotation velocities are well defined. 
We thus select the 22 galaxies from \citetalias{Verheijen01a} with flat or
declining rotation curves. 
For galaxies whose rotation curves rise until the last measured point, the
maximum and asymptotic rotation velocities are ill-defined, so we excluded
galaxies from \citetalias{Verheijen01a} with such rotation curves.  
We include all galaxies from the sample of massive, late-type spiral galaxies
from \citetalias{Spekkens06}, except UGC~2849, which appears to be interacting
with a companion and whose kinematics are lopsided. The remaining 7 galaxies
have well-defined rotation curves, some of which appear to be weakly declining
in the outer regions. However, given the uncertainties in the inclination of
the gas layer at large radii, they are also consistent with being entirely
flat.
Finally, we include all 19 galaxies from \citetalias{Noordermeer07b}. 
These are all early-type disk galaxies (S0 - Sab), and mostly massive
($V_{\mathrm {max}} \ga 200 \, {\mathrm {km \, s^{-1}}}$). 
Many of these systems have declining rotation curves, with large differences
between $V_{\mathrm {asymp}}$ and $V_{\mathrm {max}}$.

Our final sample consists of 48 galaxies. They are listed, together with
the photometric and kinematic data, in table~\ref{table:data}.
%%%%%%%%%%%%%%%%%%%%%%%%%%%%%%%%%%%%%%%%%%%%%%%%%%%%%%%%%%%%%%%%%%%%%%%%%%%%%%%
%                                                                             %
% BEGIN TABLE 1: PHOTOMETRIC AND KINEMATIC PROPERTIES FOR TULLY-FISHER        %
% ANALYSIS                                                                    %
% label: {table:data}                                                         %
%%%%%%%%%%%%%%%%%%%%%%%%%%%%%%%%%%%%%%%%%%%%%%%%%%%%%%%%%%%%%%%%%%%%%%%%%%%%%%%
\begin{table*}
 \begin{minipage}{14.35cm}
  \centering
   \caption[Photometric and kinematic properties used for the
    Tully-Fisher analysis]{Photometric and kinematic properties used 
    for the Tully-Fisher analysis: (1)~galaxy name; (2)~total apparent K-band
    magnitude and error, taken from \citetalias{Tully96} (Verheijen's
    galaxies) and the 2MASS galaxy catalogue (Spekkens' and Noordermeer's
    galaxies); (3)~correction for galactic foreground extinction, taken from
    \citet{Schlegel98}; (4)~correction for internal extinction, calculated
    using equation~\ref{eq:internalabsorption}; (5)~assumed distance;
    (6)~magnitude error due to distance uncertainty; (7)~resulting absolute 
    magnitude; (8)~maximum and (9)~asymptotic rotation velocities from the
    rotation curve and (10)~width of global \HI\ profile, corrected for
    instrumental broadening, random gas motions and inclination.  
    \label{table:data}}  
  
   \begin{tabular}{lr@{\hspace{0.1cm}$\pm$\hspace{0.1cm}}lcc%
       cccr@{\hspace{0.1cm}$\pm$\hspace{0.1cm}}l%
       r@{\hspace{0.1cm}$\pm$\hspace{0.1cm}}l%
       r@{\hspace{0.1cm}$\pm$\hspace{0.1cm}}l}
      
    \hline
    \multicolumn{1}{c}{galaxy} & \multicolumn{2}{c}{$m_K$} &  
    $A^{\mathrm {fg}}_{K}$ & $A^{i}_{K}$ & $D$ & 
    $\delta M_{\mathrm {dist}} \hspace{-0.15cm}$ &  
    $M_K^c$ & \multicolumn{2}{c}{$V_{\mathrm {max}}$} &
    \multicolumn{2}{c}{$V_{\mathrm {asymp}}$} & 
    \multicolumn{2}{c}{$W_{20,R}^{c,i}$} \\ 
     
     & 
    \multicolumn{4}{c}{-------------- mag --------------} & 
    Mpc & \multicolumn{2}{c}{------ mag ------} & 
    \multicolumn{6}{c}%
    {----------------- ${\mathrm {km \, s^{-1}}}$ -----------------} \\ 

    \multicolumn{1}{c}{(1)} & \multicolumn{2}{c}{(2)} &  
    \multicolumn{1}{c}{(3)} & \multicolumn{1}{c}{(4)} &
    \multicolumn{1}{c}{(5)} & (6) & (7) & \multicolumn{2}{c}{(8)} &     
    \multicolumn{2}{c}{(9)} & \multicolumn{2}{c}{(10)} \\
    \hline

    \multicolumn{14}{l}{Galaxies from \citet{Verheijen01a}} \\
    UGC 6399  & 11.09 & 0.08 & 0.07 & 0.06 & 18.6 & 0.17 & -20.33 &  88 & 5  & 88  & 5  & 172 & 2  \\    
    UGC 6446  &  11.5 & 0.08 & 0.07 & 0.02 & 18.6 & 0.17 & -19.88 &  82 & 4  & 82  & 4  & 174 & 8  \\    
    NGC 3726  &  7.96 & 0.08 & 0.07 & 0.05 & 18.6 & 0.17 & -23.45 & 162 & 9  & 162 & 9  & 330 & 9  \\    
    NGC 3729  &   8.6 & 0.08 & 0.05 & 0.03 & 18.6 & 0.17 & -22.78 & 151 & 11 & 151 & 11 & 295 & 14 \\    
    UGC 6667  & 10.81 & 0.08 & 0.07 & 0.10 & 18.6 & 0.17 & -20.65 &  86 & 3  & 86  & 3  & 167 & 2  \\    
    NGC 3877  &  7.75 & 0.08 & 0.10 & 0.15 & 18.6 & 0.17 & -23.76 & 167 & 11 & 167 & 11 & 335 & 6  \\    
    NGC 3917  &  9.08 & 0.08 & 0.09 & 0.12 & 18.6 & 0.17 & -22.40 & 135 & 3  & 135 & 3  & 275 & 3  \\    
    NGC 3949  &  8.43 & 0.08 & 0.09 & 0.05 & 18.6 & 0.17 & -22.98 & 164 & 7  & 164 & 7  & 320 & 8  \\    
    NGC 3953  &  7.03 & 0.08 & 0.13 & 0.08 & 18.6 & 0.17 & -24.41 & 223 & 5  & 223 & 5  & 446 & 5  \\    
    UGC 6917  &  10.3 & 0.08 & 0.12 & 0.04 & 18.6 & 0.17 & -21.10 & 104 & 4  & 104 & 4  & 224 & 7  \\    
    NGC 3992  &  7.23 & 0.08 & 0.13 & 0.08 & 18.6 & 0.17 & -24.21 & 272 & 6  & 242 & 5  & 547 & 13 \\    
    NGC 4013  &  7.68 & 0.08 & 0.07 & 0.15 & 18.6 & 0.17 & -23.83 & 195 & 3  & 177 & 6  & 377 & 1  \\    
    NGC 4010  &  9.22 & 0.08 & 0.11 & 0.17 & 18.6 & 0.17 & -22.31 & 128 & 9  & 128 & 9  & 254 & 1  \\    
    UGC 6983  & 10.52 & 0.08 & 0.12 & 0.03 & 18.6 & 0.17 & -20.87 & 107 & 7  & 107 & 7  & 222 & 4  \\    
    NGC 4085  &   9.2 & 0.08 & 0.08 & 0.11 & 18.6 & 0.17 & -22.27 & 134 & 6  & 134 & 6  & 247 & 7  \\    
    NGC 4088  &  7.46 & 0.08 & 0.09 & 0.10 & 18.6 & 0.17 & -24.00 & 173 & 14 & 173 & 14 & 362 & 5  \\    
    NGC 4100  &  8.02 & 0.08 & 0.10 & 0.14 & 18.6 & 0.17 & -23.48 & 195 & 7  & 164 & 13 & 386 & 5  \\    
    NGC 4102  &  7.86 & 0.08 & 0.09 & 0.07 & 18.6 & 0.17 & -23.57 & 178 & 11 & 178 & 11 & 393 & 10 \\   
    NGC 4138  &  8.19 & 0.08 & 0.06 & 0.05 & 18.6 & 0.17 & -23.22 & 195 & 7  & 147 & 12 & 374 & 16 \\    
    NGC 4157  &  7.52 & 0.08 & 0.09 & 0.20 & 18.6 & 0.17 & -24.04 & 201 & 7  & 185 & 10 & 398 & 4  \\    
    NGC 4183  &  9.76 & 0.08 & 0.06 & 0.14 & 18.6 & 0.17 & -21.74 & 115 & 6  & 109 & 4  & 228 & 2  \\    
    NGC 4217  &  7.61 & 0.08 & 0.08 & 0.15 & 18.6 & 0.17 & -23.90 & 191 & 6  & 178 & 5  & 381 & 5  \\[0.14cm]
    \multicolumn{14}{l}{Galaxies from \citet{Spekkens06}} \\      
    NGC 1324  &  9.15 & 0.11 & 0.02 & 0.18 & 74.2 & 0.08 & -25.40 & 305 & 7  & 275 & 9  & 604 & 15 \\    
    ESO 563G21&  8.89 & 0.11 & 0.08 & 0.22 & 59.2 & 0.10 & -25.27 & 310 & 11 & 300 & 6  & 655 & 12 \\    
    NGC 2862  &  9.27 & 0.11 & 0.01 & 0.22 & 55.9 & 0.10 & -24.70 & 280 & 5  & 250 & 10 & 574 & 11 \\    
    NGC 2955  & 10.04 & 0.11 & 0.00 & 0.08 & 95.3 & 0.06 & -24.93 & 265 & 19 & 230 & 17 & 546 & 34 \\    
    IC 4202   & 10.41 & 0.11 & 0.01 & 0.23 & 97.2 & 0.06 & -24.76 & 245 & 3  & 245 & 3  & 517 & 10 \\    
    NGC 6195  & 10.18 & 0.11 & 0.01 & 0.09 &123.3 & 0.05 & -25.38 & 255 & 17 & 255 & 17 & 536 & 37 \\    
    UGC 11455 &  9.23 & 0.11 & 0.07 & 0.26 & 75.7 & 0.08 & -25.50 & 280 & 9  & 270 & 10 & 602 & 10 \\[0.14cm] 
    \multicolumn{14}{l}{Galaxies from \citet{Noordermeer07b}} \\  
    UGC 624   &  9.26 & 0.11 & 0.02 & 0.11 & 65.1 & 0.09 & -24.94 & 300 & 16 & 270 & 30 & 599 & 38 \\
    UGC 2487  &  8.64 & 0.11 & 0.07 & 0.04 & 67.4 & 0.09 & -25.61 & 390 & 21 & 330 & 24 & 755 & 36 \\
    UGC 2916  & 10.01 & 0.11 & 0.10 & 0.02 & 63.5 & 0.09 & -24.13 & 220 & 17 & 180 & 17 & 501 & 39 \\
    UGC 2953  &  6.04 & 0.11 & 0.15 & 0.06 & 15.1 & 0.38 & -25.07 & 315 & 9  & 260 & 16 & 603 & 18 \\ 
    UGC 3205  &  9.50 & 0.11 & 0.20 & 0.11 & 48.7 & 0.12 & -24.24 & 240 & 7  & 210 & 5  & 451 & 7  \\
    UGC 3546  &  8.48 & 0.11 & 0.03 & 0.05 & 27.3 & 0.21 & -23.78 & 260 & 65 & 190 & 12 & 421 & 13 \\
    UGC 3580  & 10.12 & 0.11 & 0.02 & 0.05 & 19.2 & 0.30 & -21.37 & 127 & 5  & 125 & 5  & 246 & 7  \\
    UGC 3993  & 10.33 & 0.11 & 0.02 & 0.01 & 61.9 & 0.09 & -23.66 & 300 & 45 & 250 & 40 & 550 & 79 \\
    UGC 4458  &  9.31 & 0.11 & 0.01 & 0.02 & 64.2 & 0.09 & -24.76 & 490 & 75 & 240 & 43 & 620 & 93 \\
    UGC 4605  &  8.46 & 0.11 & 0.02 & 0.14 & 20.9 & 0.28 & -23.30 & 225 & 5  & 185 & 4  & 412 & 2  \\
    UGC 5253  &  7.36 & 0.11 & 0.01 & 0.03 & 21.1 & 0.27 & -24.30 & 255 & 20 & 210 & 21 & 495 & 34 \\
    UGC 6786  &  8.70 & 0.11 & 0.01 & 0.11 & 25.9 & 0.22 & -23.48 & 230 & 5  & 215 & 10 & 456 & 6  \\
    UGC 6787  &  7.66 & 0.11 & 0.01 & 0.08 & 18.9 & 0.31 & -23.82 & 270 & 10 & 250 & 7  & 495 & 7  \\
    UGC 8699  &  9.74 & 0.11 & 0.00 & 0.13 & 36.7 & 0.16 & -23.22 & 205 & 6  & 180 & 9  & 382 & 10 \\
    UGC 9133  &  8.80 & 0.11 & 0.01 & 0.06 & 54.3 & 0.11 & -24.94 & 300 & 15 & 225 & 12 & 537 & 18 \\
    UGC 11670 &  7.72 & 0.11 & 0.08 & 0.09 & 12.7 & 0.46 & -22.97 & 190 & 5  & 160 & 6  & 341 & 6  \\
    UGC 11852 & 10.69 & 0.11 & 0.03 & 0.05 & 80.0 & 0.07 & -23.90 & 220 & 16 & 165 & 11 & 401 & 29 \\
    UGC 11914 &  6.83 & 0.11 & 0.03 & 0.01 & 14.9 & 0.39 & -24.08 & 305 & 43 &%
                                                       \multicolumn{2}{c}{--$^\dagger$} & 583 & 85 \\
    UGC 12043 & 10.81 & 0.11 & 0.02 & 0.05 & 15.4 & 0.38 & -20.21 & 93  & 6  & 90  & 4  & 185 & 3  \\
   \hline
   \multicolumn{14}{l}{\parbox{13.95cm}{\footnotesize $^\dagger$The rotation
    curve of UGC~11914 only extends out to about 3.3 R-band disk scale lengths
    and does not show the characteristic decline that is seen in other
    galaxies of similar type and luminosity (see \citetalias{Noordermeer07b}). 
    Since in many comparable cases, the decline in the rotation velocities
    sets in outside the optical disk only, it is well possible that the
    rotation curve in UGC~11914 would also decline at larger radii if we were
    able to measure it. The asymptotic rotation velocity $V_{\mathrm {asymp}}$
    for UGC~11914 seems therefore ill-defined and we excluded this galaxy from
    the sample for the asymptotic velocity relations.}}  
   \end{tabular}
 \end{minipage}
\end{table*}  
%%%%%%%%%%%%%%%%%%%%%%%%%%%%%%%%%%%%%%%%%%%%%%%%%%%%%%%%%%%%%%%%%%%%%%%%%%%%%%%
%                                                                             %
% END TABLE 1: PHOTOMETRIC AND KINEMATIC PROPERTIES FOR TULLY-FISHER ANALYSIS %
%                                                                             %
%%%%%%%%%%%%%%%%%%%%%%%%%%%%%%%%%%%%%%%%%%%%%%%%%%%%%%%%%%%%%%%%%%%%%%%%%%%%%%%

\subsection{Photometric data}
\label{subsec:photdata}
We have chosen to study the Tully-Fisher relation using K-band
photometry. 
Near-infrared photometry has a number of clear advantages over conventional
optical data.  
Foremost, the effects of dust extinction is strongly reduced. 
Several galaxies in our sample are highly inclined with respect to the line of 
sight, such that a large fraction of the light is absorbed by internal
dust. 
In the optical bands, this effect can be very significant ($\ga 1 {\mathrm
{mag}}$), leading to large uncertainties in the derived absolute
luminosities. 
In the near infrared, these uncertainties are less severe.  
In addition, recent bursts of star formation can have a large effect on the
optical luminosity of a galaxy, whereas the K-band is much less sensitive to
such events. 

Deep K-band photometry for the galaxies from \citetalias{Verheijen01a} is
available from \citet[][hereafter T96]{Tully96}.
Neither \citetalias{Spekkens06} nor \citetalias{Noordermeer07b} had private
near-infrared photometry at their disposal, but all galaxies from these
subsamples have been observed in K in the framework of the Two Micron All Sky
Survey\footnote{2MASS (http://www.ipac.caltech.edu/2mass/) is a joint project
  of the University of Massachusetts and the Infrared Processing 
  and Analysis Center/California Institute of Technology, funded by the
  National Aeronautics and Space Administration and the National Science 
  Foundation.}.  
Although 2MASS is a very homogeneous survey, it is also less deep than the
data from \citetalias{Tully96}. 
To judge the reliability of the 2MASS photometric data, and to ensure that
we do not introduce systematic differences between our various subsamples, we
compared the K-band magnitudes from \citetalias{Tully96} with the total
extrapolated magnitudes from 2MASS, for all systems which were detected in
both surveys. 

\begin{figure}
 \centerline{\psfig{figure=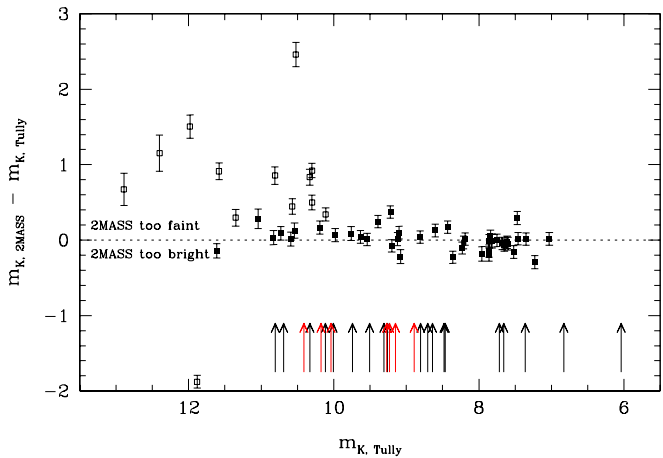,height=6.cm}}
  \caption{Comparison of 2MASS K-band magnitudes with the data from
  \citetalias{Tully96}. Data points show the difference between the total 
  apparent magnitudes from 2MASS and those from \citetalias{Tully96};
  errorbars show the combined errors. Open symbols indicate low  
  surface brightness galaxies ($\mu_{0,K} > 18 \magasas$), filled
  symbols indicate HSB systems. Black and red vertical arrows indicate the
  2MASS apparent magnitudes for the galaxies from \citetalias{Noordermeer07b}
  and \citetalias{Spekkens06} respectively.
  \label{fig:2MASScomp}}  
\end{figure}
Figure~\ref{fig:2MASScomp} shows that there is good agreement between the
2MASS photometry and Tully's data for high surface brightness (HSB) galaxies.  
For the galaxies with measured central surface brightness $\mu_{0,K} < 
18.0 \magasas$, the average difference between 2MASS' and T96's
magnitudes is 0.01~mag, with a standard deviation of 0.14~mag. 
\citetalias{Verheijen01a} reported, based on an internal comparison of 
their observations from different nights, an average photometric
uncertainty of 0.08~mag in the K-band magnitudes of
\citetalias{Tully96}.  
This implies that the uncertainties in the 2MASS magnitudes must be
approximately 0.11~mag, significantly larger than the average errors 
given by the 2MASS database (typically 0.03~mag for the HSB galaxies in
figure~\ref{fig:2MASScomp}).   
Photometric errors of 0.03~mag seem somewhat optimistic, in particular
since the 2MASS images are significantly less deep than those of
\citetalias{Tully96}.  

Figure~\ref{fig:2MASScomp} also shows that for low
surface brightness galaxies, the 2MASS magnitudes are not reliable
and sometimes deviate strongly from the values from
\citetalias{Tully96}.   
Inspection of the 2MASS images for these galaxies shows that they are
of such low surface brightness that they are often barely detected (in
fact, most of the LSB galaxies from Tully's sample are not detected at
all by 2MASS). 
It appears that 2MASS has missed a large fraction of the flux of many LSB 
galaxies, compared to the deeper data from \citetalias{Tully96}.   

The galaxies from \citetalias{Noordermeer07b} and \citetalias{Spekkens06} are
all luminous, high surface brightness galaxies. 
For these objects, it seems safe to adopt the 2MASS magnitudes, but we assume
an average photometric uncertainty of 0.11~mag instead of the smaller errors
given by 2MASS. 

To convert the apparent magnitudes to absolute luminosities, we first correct
for extinction effects. 
Corrections for Galactic foreground extinction were taken from
\citet{Schlegel98}; they are generally small ($\le 0.1$~mag for 85\% of the 
galaxies). 
Several correction schemes exist to determine the amount of {\em internal}
extinction; the most commonly used methods are based on those of
\citet{Tully85} or \citet{Giovanelli94}. 
Here, we follow \citetalias{Verheijen01a}, who employed the following 
relation, originally derived by \citet{Tully98}, for the internal
extinction parameter:
\begin{equation}
 A_\lambda^{i \rightarrow 0} = -\gamma_\lambda \log \left( \frac{b}{a}
                                                   \right),
 \label{eq:internalabsorption}
\end{equation}
with $b/a$ the observed minor-to-major axis ratio of the optical
image. 
$\gamma_\lambda$ is wavelength dependent and was found by
\citet{Tully98} to depend also on the absolute luminosity of the
galaxy: brighter galaxies contain on average more dust than fainter
ones. 
These authors used the Tully-Fisher relation itself, in an iterative
way, to express the absolute luminosity in terms of the \HI\ line
width, and give the following description for $\gamma_\lambda$:
\begin{equation}
 \gamma_\lambda = \alpha_\lambda + \beta_\lambda ( \log W_{20,R}^{c,i} -
 2.5 ),
\end{equation}
where $W_{20,R}^{c,i}$ is the \HI\ line width, corrected for
inclination and broadening due to instrumental effects and random gas
motions (see below). 
For the wavelength dependent parameters $\alpha_\lambda$ and
$\beta_\lambda$, we use the values given by \citet{Tully98}.
Note that even in the K-band used here, we find significant internal
extinction corrections, although they do not exceed 0.1~magnitude except in
the most inclined systems.  

Distances for our galaxies were derived as follows. 
All galaxies in \citetalias{Verheijen01a} belong to the Ursa Major cluster and
have therefore roughly equal distances; throughout this study, we adopt
Verheijen's value of 18.6~Mpc (taken from \citealt{Tully00}). 
For the galaxies from \citetalias{Noordermeer07b}, we used the distances as
derived by \citet{Noordermeer05}, which were based on a simple Hubble-flow
model, corrected for Virgo-centric infall and assuming a Hubble constant of
$75 \, {\mathrm {km \, s^{-1} \, Mpc^{-1}}}$. 
For the galaxies from \citetalias{Spekkens06}, we used the same
method\footnote{Note that our distance estimates for the galaxies from
  \citetalias{Spekkens06} are slightly different from those given in the
  original paper. Part of the discrepancy comes from the fact that we assume
  $h = 0.75$ instead of 0.70. Further differences arise from the treatment of
  Virgo-centric streaming motions, but are small ($<$~0.1~mag in all
  cases).}. 

In table~\ref{table:data}, we give for each galaxy the raw apparent magnitudes
and photometric errors, the galactic foreground and internal extinction
parameters, $A^{\mathrm {fg}}_K$ and $A^{i}_K$, the assumed distances and the
resulting corrected absolute magnitudes used in the remainder of this paper.  

Finally, in addition to the photometric errors, we also account for the
uncertainties in the absolute magnitudes which arise from errors in our
distance estimates.  
For the galaxies from \citetalias{Verheijen01a}, these are defined by the
depth of the Ursa Major cluster; here, we adopt Verheijen's estimate of
0.17~mag. 
For the other galaxies, we estimate the distance uncertainties by assuming a
typical peculiar velocity of $200 \, {\mathrm {km \, s^{-1}}}$. 
The resulting errors on the absolute magnitude $\delta M_{\mathrm {dist}}$
are given in column (7) of table~\ref{table:data} and were added quadratically
to the photometric errors to obtain the total uncertainties used for the
subsequent analysis.  

Note that we have not included the uncertainties in the internal extinction 
corrections in our error budget here.   
Especially for the highly-inclined systems in our samples, the corrections
$A^{i}_{\lambda}$ are significant and galaxy-to-galaxy variations of
$\gamma_\lambda$ can cause substantial deviations from the corrections used
here.  
However, without additional information about the dust content of our
galaxies, the corrections listed in table~\ref{table:data} 
are the best possible estimate and the uncertainties are very
difficult to quantify. 
We will ignore these uncertainties for the subsequent analysis here,
but note that some of the scatter in our TF relations might
be explained by the uncertainties in the internal extinction. 
\begin{figure*}
 \centerline{\psfig{figure=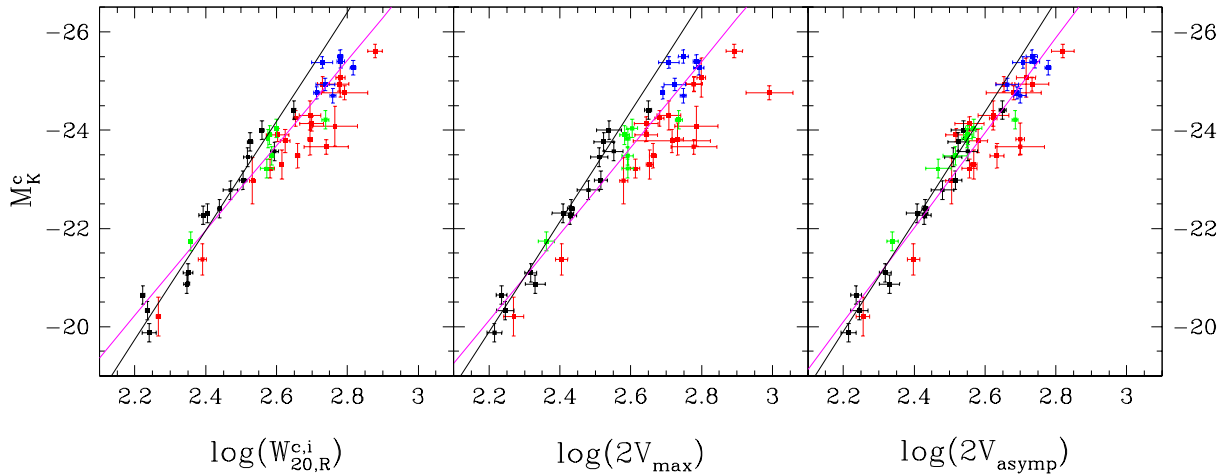,height=6.25cm}}
  \caption{Tully-Fisher relations in the K-band, using the corrected widths of
   the global \HI\ profiles ({\em left}) and the maximum ({\em middle}) and
   asymptotic rotation velocities ({\em right}). 
   Black and green data points show galaxies from \citetalias{Verheijen01a}
   with flat and declining rotation curves respectively; blue data points show
   galaxies from \citetalias{Spekkens06} and red points show galaxies from
   \citetalias{Noordermeer07b}. The black lines show the fits from
   \citetalias{Verheijen01a} to galaxies with flat rotation curves, while the
   magenta lines show the fits to the combined sample (see
   table~\ref{table:fits}).\label{fig:TF}}    
\end{figure*}

\subsection{Kinematic data}
\label{subsec:kindata}
We use three kinematic parameters for our Tully-Fisher analysis. 
The first is the inclination corrected width of the \HI\ line
profile. 
For this, we used the instrumental broadening corrected measurements
from \citet{Verheijen01b}, \citetalias{Spekkens06} and \citet{Noordermeer05}
respectively. 
We apply an additional correction for broadening of the profiles caused by
random gas motions in the gas disks, using the prescriptions given by 
\citet[][based on \citealt{Tully85}]{Verheijen01b}:  
\begin{eqnarray}
 \lefteqn{ 
   \left( W_{20,R}^{c} \right)^2 = \left( W_{20}^{c} \right)^2 \: + \:
   W_t^2 \left[ 1 - 2 e^{ - \left( \frac{W_{20}^c}{W_c} \right)^2 } \right]   
   \: - 
         } \nonumber \\ 
 & & 2 W_{20}^{c} W_t 
     \left[ 1 -   e^{ - \left( \frac{W_{20}^c}{W_c} \right)^2 } \right],
 \label{eq:profilebroadening}
\end{eqnarray}
with $W_{20}^{c}$ the profile width corrected for instrumental broadening.
$W_c$ indicates the profile width where the transition from a Gaussian
to a boxy shape occurs; we assume here $W_c = 120 \, {\mathrm {km \, s^{-1}}}$
\citep{DeVaucouleurs83}.  
In practice, most of our galaxies have profile widths larger than $W_c$,
such that equation~\ref{eq:profilebroadening} yields a linear 
subtraction of $W_t$, which gives the amount by which a profile is
broadened due to the random gas motions. 
\citet{Verheijen01b} present an extensive discussion about the
suitable choice for $W_t$; here we simply copy their preferred value
of $W_t = 22 \, {\mathrm {km \, s^{-1}}}$.
The derived profile widths $W_{20,R}^{c}$ were corrected for
inclination using the values derived in the original papers; for warped
galaxies we use the inclination in the inner regions where most of the gas is
concentrated.  

The two other kinematic parameters are the maximum rotation velocity
$V_{\mathrm {max}}$ and the asymptotic rotation velocity $V_{\mathrm
{asymp}}$ ($V_{\mathrm {flat}}$ in \citetalias{Verheijen01a}), which are both 
derived directly from the rotation curves. 

For the galaxies from \citetalias{Verheijen01a}, we copied the errors on the
kinematic parameters without further modification. 
For the galaxies from \citetalias{Noordermeer07b}, we used the inclination
uncertainty derived in the original paper to estimate the errors on the
profile widths. 
The errors on the two other kinematic parameters were estimated by eye, based
on the errors in the rotation curves.  
The latter include contributions from measurement errors, kinematic
asymmetries and inclination uncertainties; for the more face-on galaxies, the
latter are usually dominant.  
Finally, for the galaxies from \citetalias{Spekkens06}, we used the
uncertainty in the ellipticity of the optical images, given in the original
paper, to estimate the inclination uncertainty, but adopted a minimum of
$5^\circ$ to account for the possibility of undetected warps in the outer
parts of the gas layer. 
The errors on $V_{\mathrm {max}}$ and $V_{\mathrm {asymp}}$ were again
estimated by eye, based on the errors in the rotation curves. 

All three kinematic parameters and corresponding errors for
each galaxy are given in table~\ref{table:data}. 
%%%%%%%%%%%%%%%%%%%%%%%%%%%%%%%%%%%%%%%%%%%%%%%%%%%%%%%%%%%%%%%%%%%%%%%%%%%%%%%
%                                                                             %
% BEGIN TABLE 2: RESULTS FROM THE TULLY-FISHER FITS                           %
% label: {table:fits}                                                         %
%%%%%%%%%%%%%%%%%%%%%%%%%%%%%%%%%%%%%%%%%%%%%%%%%%%%%%%%%%%%%%%%%%%%%%%%%%%%%%%
\begin{table*}
 \begin{minipage}{14.25cm}
  \centering
   \caption[Results from the least-$\chi^2$ fits to the Tully-Fisher
    relations]{Results from the least-$\chi^2$ fits to the K-band Tully-Fisher 
    relations shown in figure~\ref{fig:TF}. \label{table:fits}}

   \begin{tabular}{ccr@{\hspace{0.1cm}$\pm$\hspace{0.1cm}}l%
                   r@{\hspace{0.1cm}$\pm$\hspace{0.1cm}}lccl}
    \hline  
    \multirow{3}{*}{\parbox{1.5cm}{kinematic \\ parameter}} & \# of & 
    \multicolumn{2}{c}{$M_{2.6}$} & \multicolumn{2}{c}{$S$} &
    scatter & $\chi^2_{\mathrm {red}}$ & \multicolumn{1}{c}{\hspace{-0.5cm} Q}
    \\ 

     & points & \multicolumn{2}{c}{} & \multicolumn{2}{c}{} & & & \\     

     & & \multicolumn{2}{c}{mag} & \multicolumn{2}{c}{} & mag & & \\   
    
    \hline     
    $W_{20,R}^{c,i}$      & 48 & -23.69 & 0.03 & -8.65 & 0.19 & 0.38 & 2.89 & $3.3 \cdot 10^{-11}$ \\
    $V_{\mathrm {max}}$   & 48 & -23.64 & 0.04 & -8.77 & 0.22 & 0.44 & 2.98 & $7.2 \cdot 10^{-12}$ \\
    $V_{\mathrm {asymp}}$ & 47 & -23.95 & 0.04 & -9.64 & 0.25 & 0.36 & 1.76 & $5.4 \cdot 10^{-4}$  \\
    $V_{\mathrm {asymp}}$ {\footnotesize{(UGC~3993 and 6787 excluded)}} &%
                            45 & -23.97 & 0.04 & -9.71 & 0.26 & 0.34 & 1.54 & $6.8 \cdot 10^{-3}$  \\
    \hline
   \end{tabular}
 \end{minipage}
\end{table*}  
%%%%%%%%%%%%%%%%%%%%%%%%%%%%%%%%%%%%%%%%%%%%%%%%%%%%%%%%%%%%%%%%%%%%%%%%%%%%%%%
%                                                                             %
% END TABLE 2: RESULTS FROM THE TULLY-FISHER FITS                             %
%                                                                             %
%%%%%%%%%%%%%%%%%%%%%%%%%%%%%%%%%%%%%%%%%%%%%%%%%%%%%%%%%%%%%%%%%%%%%%%%%%%%%%%

%%%%%%%%%%%%%%%%%%%%%%%%%%%%%%%%%%%%%%%%%%%%%%%%%%%%%%%%%%%%%%%%%%%%%%%%%%%%%%%
%                                                                             %
%  3. Tully-Fisher relations                                                  %
%  \label{sec:TFrelations}                                                    %
%                                                                             %
%%%%%%%%%%%%%%%%%%%%%%%%%%%%%%%%%%%%%%%%%%%%%%%%%%%%%%%%%%%%%%%%%%%%%%%%%%%%%%%
\section{Tully-Fisher relations}
\label{sec:TFrelations}
In figure~\ref{fig:TF}, we show the Tully-Fisher relations for the combined
sample of galaxies. 
Compared to the study in \citetalias{Verheijen01a}, our combined sample
greatly increases the number of luminous galaxies, with our most luminous
system, UGC~2487, being 1.2 magnitudes brighter than the most luminous
object in Verheijen's sample.   

We fitted direct linear relations of the following form to the combined data
sets: 
\begin{equation}
M_K^c = M_{2.6} + S \left( \log W - 2.6 \right),
\end{equation}
where $M_K^c$ is the extinction corrected absolute magnitude and W is either
$W_{20,R}^{c,i}$, $2 V_{\mathrm {max}}$ or $2 V_{\mathrm {asymp}}$. 
We calculated the zeropoint $M_{2.6}$ at $\log W = 2.6$, rather than $0$, to
avoid strong covariances between the two free parameters. 
The fits were done using a simple $\chi^2$-minimisation procedure, taking the
errors in both directions into account. 
They are indicated with the magenta lines in figure~\ref{fig:TF}; for
comparison, we also show (in black) the fits to the `F'-sample from
\citetalias{Verheijen01a}, i.e.\ the fits made to galaxies with flat rotation
curves only.  

The results of the fits to the combined sample are summarised in
table~\ref{table:fits}. 
The scatter around each of the fits given in this table is a weighted
rms scatter,  with the weight for each data point given by $w_i =
(\sigma_{M,i}^2 + S^2 \sigma_{\log W,i}^2)^{-1}$. 
Here, $\sigma_{M,i}$ are the magnitude errors, $\sigma_{\log W,i}$ are
the errors in the parameters on the $x$-axes and $S$ are the 
fitted slopes of the relations. 

A number of interesting results can be recognised from
figure~\ref{fig:TF} and table~\ref{table:fits}. 
First of all, our data strongly suggest a `kink' in the Tully-Fisher
relation: the TF relation seems to become shallower above a rotation
velocity of about $200 \, {\mathrm {km \, s^{-1}}}$ (equivalent to $M_K^c
\approx -23.75$).  
This is consistent with the claim first made by \citet{Peletier93},
but the effect is much more clearly visible here than in their data.     
The kink is most apparent when using the width of the global profile
($W_{20,R}^{c,i}$) or the maximum rotation velocity $V_{\mathrm
{max}}$ as kinematic parameter. 
In the left and middle panels in the figure, almost all galaxies from
the \citetalias{Spekkens06} and \citetalias{Noordermeer07b} samples lie to the
right of the relation defined by the intermediate-mass galaxies with flat
rotation curves from \citetalias{Verheijen01a}. 
As a result, the fits to the combined samples have a shallower slope
than those from \citeauthor{Verheijen01a} to the galaxies with flat rotation
curves.    
In particular, many of the massive, early-type disk galaxies from
\citetalias{Noordermeer07b} lie far away from the relation defined by the
later-type spirals from \citeauthor{Verheijen01a} and \citetalias{Spekkens06};
when expressed in terms of a luminosity offset, their distance from the main
relation can be as large as 2 magnitudes. 

However, the turnover is greatly reduced when using the asymptotic rotation
velocity $V_{\mathrm {asymp}}$ as kinematic parameter.  
Many of the galaxies in our sample with $V_{\mathrm{max}} > 200 \, {\mathrm
{km \, s^{-1}}}$, in particular the early-type disks from
\citetalias{Noordermeer07b}, have strongly declining rotation curves. 
Galaxies such as UGC~4458, 4605, 3546, which lie far to the right of the main
relation in the left and middle panels, shift to the left when the (lower)
asymptotic velocity is used, thereby straightening the TF relation.  
This effect is also reflected in the values for the rms scatter and $\chi^2$
of the data around the fits and the `goodness-of-fit' parameters $Q$. 
Table~\ref{table:fits} shows that the TF relations using $V_{\mathrm {asymp}}$ 
are much better represented by straight lines than the relations with the
other two parameters.  

It is important, however, to note that the kink does not disappear completely
when using $V_{\mathrm {asymp}}$ as kinematic parameter. 
Even in the right hand panel of figure~\ref{fig:TF}, most galaxies at the
bright end lie to the right of the relation defined by the less luminous
galaxies with flat rotation curves from \citetalias{Verheijen01a}, indicating
that the slope of the TF relation also changes when the asymptotic rotation
velocities $V_{\mathrm {asymp}}$ are used.  
This is also confirmed when we fit linear relations to the subsamples of
galaxies fainter and brighter than $M_K^c \approx -24$; in this case, we
find that the slope changes significantly, from $-10.5 \pm 0.5$ to $-7.7 \pm
0.9$ respectively. 

The scatter in our relation using the asymptotic rotation velocities
is systematically larger than the corresponding value found by
\citetalias{Verheijen01a}.  
This difference can partly be attributed to the fact that Verheijen's
galaxies all lie in the Ursa-Major cluster, such that the distance
uncertainties are small (at least in a relative sense). 
The galaxies from \citetalias{Spekkens06} and \citetalias{Noordermeer07b} lie
predominantly in the field, where peculiar motions with respect to the Hubble
flow lead to errors in the derived distances and absolute luminosities (see
table~\ref{table:data}), and thus to additional scatter in the TF relations. 
More importantly however, the scatter in our relations is artificially 
increased by the deviations from a straight line. 
Even in the relations using the asymptotic rotation velocity
$V_{\mathrm {asymp}}$, the kink around $200 \, {\mathrm {km \, s^{-1}}}$
causes systematic deviations from the fitted straight lines. 
Since \citetalias{Verheijen01a} had only 2 galaxies with $V_{\mathrm
{asymp}} > 200 \, {\mathrm {km \, s^{-1}}}$, the change in slope at the high
mass end had virtually no influence on the scatter around the fits for his
data.  
 
\begin{figure}
 \centerline{\psfig{figure=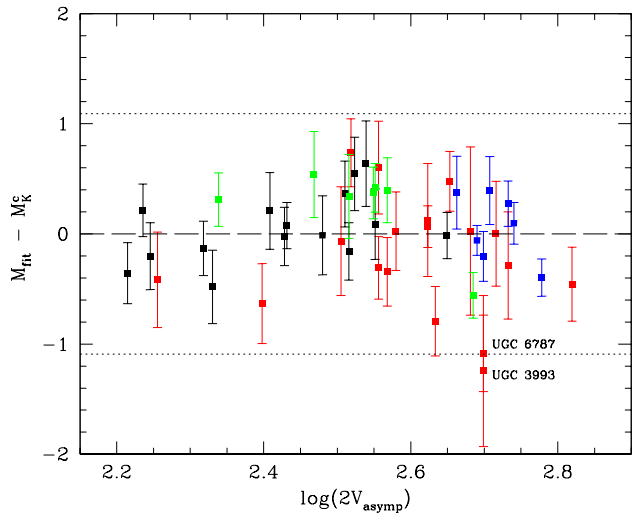,height=6.75cm}}
 \caption{Deviations of our objects with respect to the K-band vs.\
  $V_{\mathrm {asymp}}$ TF relation from table~\ref{table:fits}. Symbol
  colours are the same as in figure~\ref{fig:TF}. Errorbars are
  the effective errors, calculated by combining the magnitude and velocity
  uncertainties (see text). Dotted lines indicate the formal $3 \sigma$
  limits. \label{fig:residuals}}
\end{figure}

Finally, the scatter in the relations with $V_{\mathrm {asymp}}$ is
heavily influenced by two galaxies with unusually large deviations
from the main relation: UGC~3993 and UGC~6787 ($3.4$ and $3.0 \sigma$
outliers respectively).   
In figure~\ref{fig:residuals}, we show the deviations of the
galaxies from our sample with respect to the $M_K^c$ vs.\ $V_{\mathrm
{asymp}}$ fit described in table~\ref{table:fits}.  
The errorbars in the figure take the uncertainties of the points in
both directions into account, and were calculated as 
$( \Delta \, (M - M_{\mathrm {fit}}) )^2 = (\Delta M)^2 + (\Delta \log(2V)
\cdot S)^2$, with $S = -9.64$ the slope of the fit.

UGC~3993 is a nearly face-on galaxy ($i \approx 20\deg$, see
\citetalias{Noordermeer07b}), with correspondingly large uncertainties in the
derived rotational velocities (also reflected in the large errorbars in
figures~\ref{fig:TF} and \ref{fig:residuals}).  
The deviation of this galaxy may be due partly to a slight under-estimation of
the true inclination angle.
It seems unlikely, however, that the inclination uncertainty is responsible
for the full deviation of UGC~3993; to bring this galaxy to the middle of the
relation would require an inclination angle of 30\deg, a value which seems to
fall outside the range supported by the observational data (see
\citetalias{Noordermeer07b}).

The most likely explanation for the offset of UGC~6787 from the main relation
lies in the distance determination.  
UGC~6787 lies close to the center of the Ursa~Major cluster, but has a
redshift that is about $200 \, {\mathrm {km \, s^{-1}}}$ higher than the high 
velocity envelope of the cluster \citep{Tully96}. 
It may thus lie behind the main cluster, which might in turn imply that it is
being drawn into the cluster and that the recession velocity is lower than
expected in the case of pure Hubble flow.  
In that case, our adopted distance of 18.9~Mpc and the derived luminosities
are too small, explaining the offset in the TF relations.    

For completeness, we also list in table~\ref{table:fits} the
results of fits to the TF relations with the asymptotic rotation
velocities when UGC~3993 and 6787 are excluded from the sample. 
It is clear that the exclusion of the two points from the fits does
not lead to significantly different slopes or zeropoints.
As expected, the scatter is reduced, but since the kink in the
relations has not been removed by the exclusion of the two discrepant
points, the values are still larger than the corresponding ones from
\citetalias{Verheijen01a} and the $\chi^2$ and $Q$-parameter indicate
that the deviations from a straight relation are still real.

%%%%%%%%%%%%%%%%%%%%%%%%%%%%%%%%%%%%%%%%%%%%%%%%%%%%%%%%%%%%%%%%%%%%%%%%%%%%%%%
%                                                                             %
%  4. The Baryonic Tully-Fisher relation                                      %
%  \label{sec:BaryonicTF}                                                     %
%                                                                             %
%%%%%%%%%%%%%%%%%%%%%%%%%%%%%%%%%%%%%%%%%%%%%%%%%%%%%%%%%%%%%%%%%%%%%%%%%%%%%%%
\section{Discussion}
\label{sec:discussion}

\subsection{Systematic variations along the TF relation}
\label{subsec:systematics}
What could be the origin of the change of slope in the Tully-Fisher
relation? Does it truly indicate a break in the relation between
baryons and dark matter, or could it be explained by other effects? 
An obvious possibility is that some of the corrections from
sections~\ref{subsec:photdata} and \ref{subsec:kindata} are inaccurate for the
brightest galaxies. Such an effect may lead to a systematic deviation of these
galaxies from the TF relation for lower-mass galaxies, precisely as observed. 
\begin{figure}
 \centerline{\psfig{figure=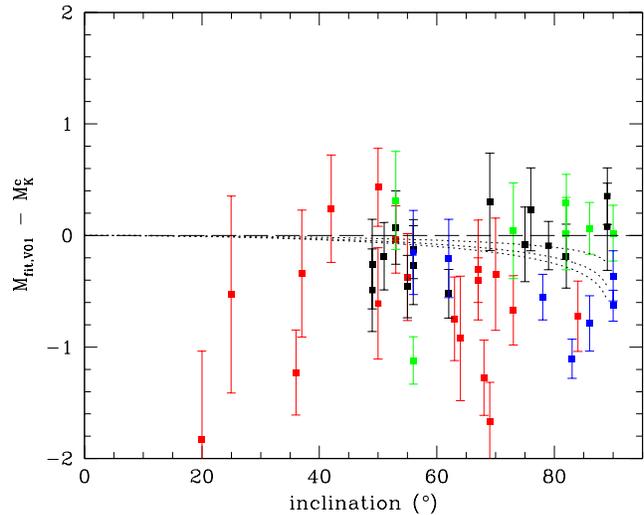,height=6.75cm}}
 \caption{Deviations of our objects with respect to Verheijen's K-band vs.\
  $V_{\mathrm {asymp}}$ TF relation for galaxies with flat rotation curves, as
  a function of inclination. Symbol colours are the same as in
  figure~\ref{fig:TF}. The dotted lines show the adopted internal extinction
  corrrections for fiducial galaxies with \HI\ profile widths of 200, 400 and
  $600 \, {\mathrm {km \, s^{-1}}}$ (top to bottom respectively). 
  \label{fig:Vresiduals_vs_i}}  
\end{figure} 

\subsubsection{internal extinction}
The largest uncertainty in the corrections which we applied to our raw data
lies, most probably, in the treatment of the internal extinction. 
We corrected for internal extinction following \citet{Tully98}.
They derived their corrections from a sample of galaxies which spans a
roughly similar luminosity range as ours and, importantly, observed no
systematic deviations of the brightest galaxies. 
On the other hand, \citet{Giovanelli95} {\em did} find a hint for excess
extinction in the brightest spirals.
Given the uncertainties, it is useful to explicitly verify whether the
deviations of our most massive galaxies with respect to the TF relation for
low- and intermediate-mass galaxies can be explained as being due to an
underestimation of the internal extinction in these systems. 
This is done in figure~\ref{fig:Vresiduals_vs_i}, where we plot the deviations
of all galaxies with respect to \citeauthor{Verheijen01a}'s TF relation for
spirals with flat rotation curves, as a function of inclination. 

Figure~\ref{fig:Vresiduals_vs_i} shows a tentative trend for the galaxies from
\citetalias{Spekkens06} (blue points), with the edge-on systems deviating more
than the galaxies with intermediate inclinations. 
To explain this trend, about a factor three more extinction is required in
these galaxies than assumed in section~\ref{subsec:photdata}.  
The galaxies from \citetalias{Noordermeer07b} (red points), however, do not
follow an obvious trend and strongly deviating cases are found among both
near-face-on and highly-inclined galaxies. 

Based on our data, we cannot strictly rule out the possibility that the
internal extinction in the most massive spiral galaxies is higher than we
assumed.  
However, figure~\ref{fig:Vresiduals_vs_i} shows that if this is indeed the
case, it still cannot explain the deviations of the low-inclination galaxies. 
It therefore seems unlikely that the change of slope at the high-mass end of
the Tully-Fisher relation can be explained purely as an effect of excess
extinction in the most massive disk galaxies. 

\subsubsection{stellar population variations}
\label{subsubsec:colmag}
\begin{figure}
 \centerline{\psfig{figure=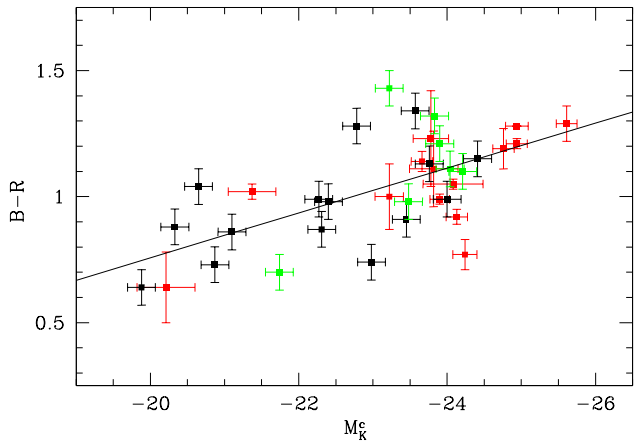,height=5.75cm}}
 \caption{Colour-magnitude relation for all galaxies in our sample for which
   B-R colours are available. Colour information for galaxies from
   \citetalias{Verheijen01a} were taken from the same paper, whereas
   \citet{Noordermeer07a} give colours for a subset of the galaxies from
   \citetalias{Noordermeer07b}. The black line shows a least-squares linear
   fit to the data points. \label{fig:colmag}}
\end{figure}

Another possibility which could explain a change of slope in the Tully-Fisher
relation is that the most massive galaxies have stellar populations with higher
mass-to-light ratios than the less luminous systems, such that they are
underluminous for a given rotation velocity. 
To first order, such an effect is likely to be present in disk galaxies, given
the well-known fact that bright galaxies tend to be redder than their
less-luminous counterparts \citep{Visvanathan81,Tully82}. 
This colour-magnitude relation is also present in our own sample: in
figure~\ref{fig:colmag}, we plot the B-R colours as a function of
absolute luminosity, for the subset of our galaxies for which the former are
available.   

The colour-magnitude relation itself cannot, however, be responsible for the
kink in the TF relation. 
Figure~\ref{fig:colmag} shows that the colour changes linearly with absolute
magnitude. 
Since mass-to-light ratio changes exponentially with colour \citep{Bell01},
this gradient will translate in a linear effect in the Tully-Fisher relation. 
In other words, a linear colour-magnitude relation will introduce a change in
the {\em global} slope of the Tully-Fisher relation.
Specifically, using the fitted slope of the colour-magnitude relation in
figure~\ref{fig:colmag}, in conjunction with the slope of the colour
vs. mass-to-light ratio relation from \citeauthor{Bell01}, it can be derived
that the stellar {\em mass} Tully-Fisher relation will be 10\% steeper than
the relations shown in figure~\ref{fig:TF} and table~\ref{table:fits}. 
Importantly, however, a linear colour-magnitude relation cannot introduce a
turnover in the TF relation.  
Only a break in the colour-magnitude relation at the same luminosity as
the kink in the TF relation can explain the latter, but
figure~\ref{fig:colmag} shows that such a feature is not present in our
sample.

\subsection{The Baryonic Tully-Fisher relation}
\label{subsec:BaryonicTF}
Having established that the observed change of slope in our K-band vs.\
$V_{\mathrm {asymp}}$ TF relation is not due to variations in internal
extinction or the properties of the stellar populations, we now consider
whether trends in relative gas content play a r\^ole. 
In this context, it is interesting to consider the so-called `Baryonic
Tully-Fisher relation', first discussed by \citet[][see also
\citealt{McGaugh05}]{McGaugh00}. 
They showed that there exists another break in the Tully-Fisher
relation at the low luminosity end (around $V_{\mathrm {rot}} =
90 \, {\mathrm {km \, s^{-1}}}$), below which galaxies are also
under-luminous.  
They were, however, able to restore a linear TF relation when, instead
of using the stellar luminosity, they adopted the total observed baryonic
mass (stars and gas). 
Since dwarf galaxies contain on average more gas than higher-luminosity
spirals \citep[relative to the optical luminosity, e.g.][]{Haynes84,
Verheijen01b, Swaters02, Geha06}, the former shifted upwards more than the
latter, and the kink in the Tully-Fisher relation disappeared.  
Luminous galaxies, on the other hand, generally contain relatively little gas.
In particular, in \citet{Noordermeer05}, it was shown that some of the
early-type disk galaxies from \citetalias{Noordermeer07b} are very gas-poor.  
In such galaxies, the baryonic budget is dominated by the stars and adding the
gas contribution will not lead to a significant increase in brightness. 

We have investigated whether the observed break at the high mass end in our
Tully-Fisher relations can be explained as a result of the relatively low
gas-content of the most luminous galaxies in our sample. 
To this purpose, we have converted the total K-band luminosities to stellar
masses, assuming an average mass-to-light ratio of the stellar populations of
0.8. 
The choice of $M_{*}/L_K = 0.8$ was made following \citet{McGaugh00},
and is consistent with the average values found from maximum-disk fits
from \citet{Verheijen97} and \citet{Palunas00}.
In reality, the values of $M_{*}/L_K$ are expected to vary from galaxy
to galaxy, but the variations will be modest in the K-band and we ignore this
variation for the moment.
The possibility that the assumption of constant M/L gives rise to additional
scatter in the baryonic TF relation is investigated below. 

\begin{figure*} 
 \centerline{\psfig{figure=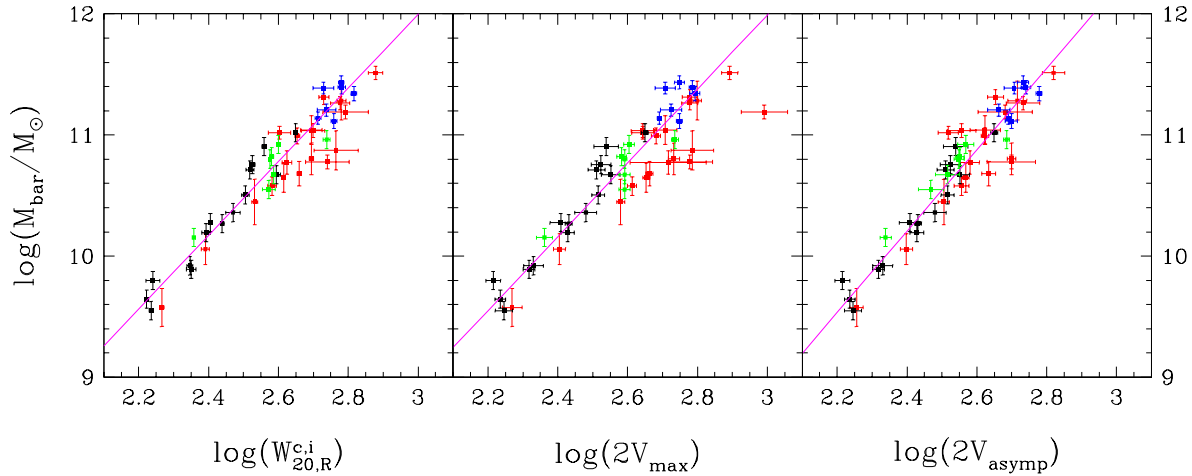,height=6.25cm}}
  \caption{Baryonic Tully-Fisher relation. The total K-band luminosity was
  converted to stellar mass, assuming a mass-to-light ratio of the stars of
  0.8 (see text), and added to the neutral gas mass. Colours of symbols and
  lines are the same as in figure~\ref{fig:TF}. \label{fig:BaryonicTF}}    
\end{figure*}

For the gas masses, we used the total \HI\ masses as given by 
\citetalias{Verheijen01a}, \citet{Noordermeer05} and \citetalias{Spekkens06}
respectively, multiplied by 1.43 to account for the presence of helium. 
The stellar and gas masses were then added to obtain an estimate for the total
baryonic mass content in our galaxies. 
The resulting baryonic TF relations are shown in figure~\ref{fig:BaryonicTF}
for the same three kinematic parameters as used before; the resulting
least-$\chi^2$ fits are summarized in table~\ref{table:BaryonicTF}.
%%%%%%%%%%%%%%%%%%%%%%%%%%%%%%%%%%%%%%%%%%%%%%%%%%%%%%%%%%%%%%%%%%%%%%%%%%%%%%%
%                                                                             %
% BEGIN TABLE 3: RESULTS FROM THE BARYONIC TULLY-FISHER FITS                  %
% label: {table:BaryonicTF}                                                   %
%%%%%%%%%%%%%%%%%%%%%%%%%%%%%%%%%%%%%%%%%%%%%%%%%%%%%%%%%%%%%%%%%%%%%%%%%%%%%%%
\begin{table*}
 \begin{minipage}{13.7cm}
  \centering
   \caption[Results from the least-$\chi^2$ fits to the Baryonic Tully-Fisher
    relations]{Results from the least-$\chi^2$ fits to the Baryonic
    Tully-Fisher relations shown in figure~\ref{fig:BaryonicTF}.  
    \label{table:BaryonicTF}} 

   \begin{tabular}{cr@{\hspace{0.1cm}$\pm$\hspace{0.1cm}}l%
                   r@{\hspace{0.1cm}$\pm$\hspace{0.1cm}}lccl}
    \hline
    \multirow{3}{*}{\parbox{1.5cm}{kinematic \\ parameter}} & 
    \multicolumn{2}{c}{$(\log M)_{2.6}$} & \multicolumn{2}{c}{S} &
    scatter & $\chi^2_{\mathrm {red}}$ & \multicolumn{1}{c}{\hspace{-0.5cm} Q}
    \\ 

     & \multicolumn{2}{c}{} & \multicolumn{2}{c}{} & & & \\     

     & \multicolumn{2}{c}{} & \multicolumn{2}{c}{} & mag$^a$ & & \\   
    
    \hline 
    
    $W_{20,R}^{c,i}$      & 10.78 & 0.01 & 3.04 & 0.08 & 0.32 & 2.17 & $2.2 \cdot 10^{-6}$ \\
    $V_{\mathrm {max}}$   & 10.77 & 0.01 & 3.05 & 0.09 & 0.40 & 2.57 & $2.4 \cdot 10^{-9}$ \\
    $V_{\mathrm {asymp}}$ & 10.88 & 0.02 & 3.36 & 0.10 & 0.36 & 1.92 & $7.7 \cdot 10^{-5}$ \\
    $V_{\mathrm {asymp}}$ {\footnotesize{(UGC~3993 and 6787 excluded)}} &%
                            10.89 & 0.02 & 3.38 & 0.10 & 0.34 & 1.72 & $1.0 \cdot 10^{-3}$ \\    
    \hline 
    \multicolumn{8}{l}{\parbox{13cm}{$^a$For consistency with
    table~\ref{table:fits}, we express the scatter in magnitudes. The scatter
    in $\log(M_{\mathrm {bar}})$, as shown in figure~\ref{fig:BaryonicTF}, is
    a factor 2.5 lower.}}
   \end{tabular}
 \end{minipage}
\end{table*}  
%%%%%%%%%%%%%%%%%%%%%%%%%%%%%%%%%%%%%%%%%%%%%%%%%%%%%%%%%%%%%%%%%%%%%%%%%%%%%%%
%                                                                             %
% END TABLE 3: RESULTS FROM THE TULLY-FISHER FITS                             %
%                                                                             %
%%%%%%%%%%%%%%%%%%%%%%%%%%%%%%%%%%%%%%%%%%%%%%%%%%%%%%%%%%%%%%%%%%%%%%%%%%%%%%%

As expected from the general trends of gas content with total luminosity
mentioned above, the baryonic Tully-Fisher relations have a shallower slope
than the standard stellar TF relations. 
But more importantly, the `kink' in the TF relations from figure~\ref{fig:TF}
is reduced. 
The relations with $W_{20,R}^{c,i}$ and $V_{\mathrm {max}}$ are better
represented by a straight line than the original relations in
figure~\ref{fig:TF}; the scatter and $\chi^2$ are reduced with
respect to the original values and the $Q$-parameters are higher. 
Thus, our data show that the concept of the Baryonic Tully-Fisher relation is
not only useful to increase the linearity in the TF relation at the low
luminosity end, but that in addition, it also reduces the kink around $200 \,
{\mathrm {km \, s^{-1}}}$.   
However, this effect is smaller than the one discussed in the previous
section, and the Baryonic Tully Fisher relations using $W_{20,R}^{c,i}$ or
$V_{\mathrm {max}}$ are still worse than our original, {\em stellar}
luminosity vs.\ $V_{\mathrm {asymp}}$ relation shown in figure~\ref{fig:TF}
and table~\ref{table:fits}. 

The inclusion of the gas does, however, remove the small kink that was still
present in the latter relation as well, and the baryonic mass vs.\ $V_{\mathrm
{asymp}}$ TF relation appears to be consistent with a linear relation over the
full extent of our data.
However, at the same time, the scatter around the mean relation does not
appear to have decreased and the formal quality of the fit, as measured with
the $\chi^2$ and $Q$-parameter, is actually reduced.
\begin{figure}
 \centerline{\psfig{figure=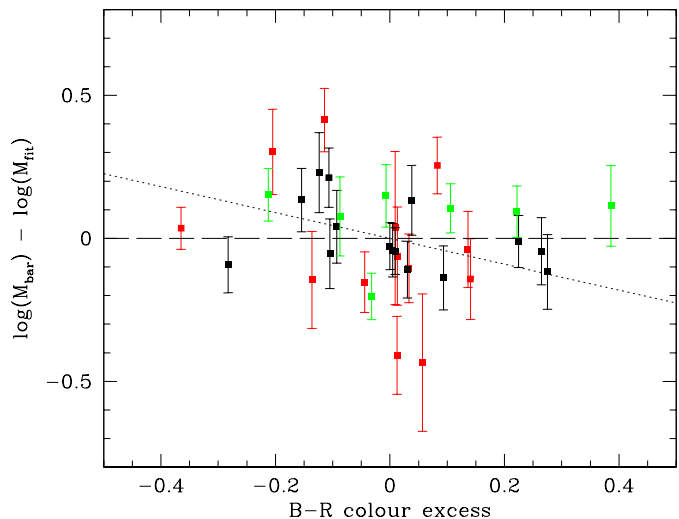,height=6.75cm}}
 \caption{Deviations of our objects with respect to the baryonic mass vs.\
  $V_{\mathrm {asymp}}$ TF relation, as a function of the colour excess of
  each galaxy with respect to the linear fit shown in figure~\ref{fig:colmag}. 
  Symbol colours are the same as in figure~\ref{fig:TF}. The dotted line shows
  the behaviour predicted by the stellar population synthesis models from
  \citet{Bell01}. \label{fig:barresiduals_vs_col}}  
\end{figure} 

The scatter might be reduced if more accurate K-band mass-to-light ratios
become available for individual galaxies instead of the constant value of
$M_{*}/L_K = 0.8$ we assumed here.
We investigate this possibility in figure~\ref{fig:barresiduals_vs_col}, where
we plot the deviation of each galaxy with respect to the baryonic mass vs.\
$V_{\mathrm {asymp}}$ TF relation, as a function of its colour excess with
respect to the linear fit shown in figure~\ref{fig:colmag}.   
As discussed in section~\ref{subsubsec:colmag}, the global colour-magnitude
trend only leads to a change in slope of the TF relation. 
The deviations from this global trend, however, are expected to give rise to
additional scatter in the TF relation, hence the use of the colour 
{\em excess} on the abscissa. 

It is clear from figure~\ref{fig:barresiduals_vs_col} that there is little
relation between the offsets from the baryonic TF relation and the colours. 
In particular, the data points do not follow the trend as predicted on the
basis of the stellar populations synthesis models from \citet{Bell01}. 
Given the uncertainties in these models and in our own data, we cannot yet
rule out the possibility that part of the scatter in the baryonic Tully-Fisher 
relation {\em is} related to variations in the mass-to-light ratios, but the
current data suggest that this is not the main contribution. 
Thus, most of the scatter in the baryonic Tully-Fisher relation must be
related to other observational effects, or be intrinsic. 

Note that the value of the $\chi^2$ parameter for the baryonic $V_{\mathrm
{asymp}}$ TF relation (1.92 or 1.72, depending on whether or not the outlying
UGCs 3993 and 6787 are included in the fit) implies that roughly 50\% of the
scatter in this relation results from the observational uncertainties in the
data points. 
This leaves an upper limit of approximately 0.25 magnitude ($\approx 25\%$
luminosity) for the intrinsic TF scatter in our galaxies. 
Given the fact that we have not included the uncertainties in the internal
extinction corrections or the mass-to-light ratios in our errorbars and that
there may well be additional uncertainties which we have ignored, it seems
safe to assume that this is a conservative upper limit and that the intrinsic
TF scatter is very small indeed.

%%%%%%%%%%%%%%%%%%%%%%%%%%%%%%%%%%%%%%%%%%%%%%%%%%%%%%%%%%%%%%%%%%%%%%%%%%%%%%%
%                                                                             %
%  5. Concluding remarks                                                      %
%  \label{sec:conclusions}                                                    %
%                                                                             %
%%%%%%%%%%%%%%%%%%%%%%%%%%%%%%%%%%%%%%%%%%%%%%%%%%%%%%%%%%%%%%%%%%%%%%%%%%%%%%%
\section{Concluding remarks}
\label{sec:conclusions}
In this study, we have used K-band photometry and high-quality rotation
curves to show that there is a `break' in the Tully-Fisher relation around a
rotation velocity of about $200 \, {\mathrm {km \, s^{-1}}}$ (equivalent to
$M_K^c \approx -23.75$, or $M_* \approx 0.7 \times 10^{11} \msun$), above
which most galaxies rotate faster than expected (or equivalent, are less
luminous).    
This kink is most pronounced in the traditional formulations of the TF
relation, using stellar luminosities and the widths of the \HI\ profiles or 
the maximum rotational velocities, and we have shown that it is mainly due to 
massive early-type disk galaxies with declining rotation curves, which lie
systematically to the right of the relation defined by less massive and
later-type spiral galaxies. 
The kink is also present, albeit to a lesser extent, in the relation with the
asymptotic rotation velocity at large radii. 
In this case, there is no systematic offset between early- and late-type
spirals, and the change in slope appears independent of morphological type. 

It is important to note that the `elliptical galaxy Tully-Fisher relation'
(using circular velocities derived from dynamical models, rather than direct
observations) shows a similar change in slope, at approximately the same
location \citep{Gerhard01, DeRijcke07}. 
Thus, the change in slope of the stellar luminosity vs.\ circular velocity
relation appears to hold universally and along the entire Hubble sequence. 
Note, however, that due to the limited spatial extent of the dynamical tracers
in elliptical galaxies, it is not generally possible to measure the asymptotic
rotation velocity in these systems. 
It can therefore not be ruled out that the kink in the elliptical galaxy TF
relation would (partly) disappear if one could use the circular velocity at
radii comparable to the outer edges of the \HI\ disks in spiral galaxies. 

A change in slope at the high-luminosity end has important consequences for
the use of the Tully-Fisher relation as a tool for estimating distances to
galaxies or for probing galaxy evolution.
In recent years, many studies have addressed the evolution of the Tully-Fisher
relation on cosmological timescales (i.e. between $z \approx 1$ and $0$),
although the results are still somewhat ambiguous. 
In their pioneering studies, \citet{Vogt96,Vogt97} reported that galaxies at
redshift $z \sim 1$ were on average 0.6~mag brighter (B-band) than galaxies in
the local universe.
Other authors find much larger luminosity evolution \citep{Milvang-Jensen03,
Bamford06}, up to 1.5 -- 2 magnitudes at redshifts of 0.25 -- 0.45
\citep{Rix97, Simard98}, whereas it has also been claimed recently that this
is all due to observational effects, and that in reality, there has been no
luminosity evolution whatsoever \citep{Flores06}. 
Given the uncertainties in the measured zeropoint offsets, it is not
surprising that measurements of evolution in the TF slope disagree as well.  
\citet{Ziegler02}, \citet{Bohm04} and \citet{Bohm06} claimed that the
evolution in the Tully-Fisher relation is luminosity dependent, with high mass
galaxies ($V_{\mathrm {max}} > 150 \, {\mathrm {km \, s^{-1}}}$) showing
little or no evolution, but low mass galaxies being up to 2 magnitudes
brighter at high redshift.  
In contrast, \citet{Weiner06} found that massive galaxies evolve more than
fainter ones. 

Whatever concensus will be reached eventually, the presence of a change of
slope in the local TF relation needs to be taken into account when
interpreting the observed evolution. 
Our results indicate that high mass galaxies are {\em under}-luminous
in the local universe, compared to a simple extrapolation of the
linear relation for lower-luminosity galaxies.
This offset is strongest for early-type disks, i.e.\ galaxies with a
high-surface brightness bulge, and for the TF relation based on the maximum
rotation velocity $V_{\mathrm {max}}$. 
Crucially, high-redshift studies will often be biased towards such
high-surface brightness galaxies. 
Moreover, when based on optical spectroscopic measurements, they will not be
able to detect a declining rotation curve at large radii, and thus be
forced to use $V_{\mathrm {max}}$, rather than $V_{\mathrm {asymp}}$. 
These two selection effects, when applied to our own data, give rise to an
offset of high-mass galaxies from the main relation, defined by low-mass
galaxies with flat rotation curves, of 1 -- 2 magnitudes (middle panel in
figure~\ref{fig:TF}).    
Thus, the luminosity evolution of high mass galaxies may be much larger than
derived by the authors mentioned above.  

We have also shown that the `kink' in the TF relation can be reduced in two
different ways, namely {\em 1}) by using the asymptotic rotation velocity from
the rotation curve as kinematic parameter and {\em 2}) by using the total
baryonic mass (stars + gas) rather than luminosity. 
The first correction appears to be the most important one for high mass
galaxies and the TF relations using the asymptotic rotation velocities show
only a small deviation from linearity.  
However, the inclusion of the gas mass also improves the linearity and only
when both refinements are used in conjunction does the `kink' disappear
completely and is a linear relation recovered.  

It is, again, interesting to note that the kink in the elliptical galaxy TF
relation also disappears when using the total baryonic mass rather than
stellar luminosity \citep{Gerhard01,DeRijcke07}, so that galaxies along the
entire Hubble sequence appear to follow the same, linear, baryonic TF
relation.  
Our results seem a strong confirmation of the idea that the baryonic
Tully-Fisher relation is fundamentally a relation between the mass of the dark
matter haloes (which define $V_{\mathrm {asymp}}$) and the total baryonic mass
in galaxies and that it holds universally, regardless of how the baryons are
distributed within the haloes \citep[cf.][]{McGaugh00, Verheijen01a}.

\section*{Acknowledgements}
We would like to thank Kristine Spekkens for kindly providing the rotation
curves for her rapidly rotating late-type spiral galaxies and for stimulating
discussions which helped to significantly improve the paper. We are also
grateful to Thijs van der Hulst, Renzo Sancisi and Pirin Erdogdu for helpful
discussions during the early stages of this project. Finally, we thank the
anonymous referee for helpful suggestions to improve the paper.   

\bibliographystyle{mn2e}
\bibliography{../../references/abbrev,../../references/refs}

\end{document}